\documentclass[letterpaper]{jpconf}
\usepackage{graphicx}
\usepackage{floatflt}
\usepackage{citesort}
\newcommand{\ra}{$\rightarrow$}
\begin{document}
\title{The crucial role of CLEO-c in the measurement of $\gamma$}

\author{C M Thomas$^{1,2}$ on behalf of the CLEO collaboration}

\address{$^1$ University of Oxford, Denys Wilkinson Building, Oxford, OX1~3RH, UK}
\address{$^2$ Science and Technology Facilities Council, Rutherford Appleton Laboratory, Harwell Science and Innovation Campus, Didcot, OX11~0QX, UK}

\ead{c.thomas2@physics.ox.ac.uk}

\begin{abstract}
The most sensitive method to measure the CKM angle $\gamma$ is to exploit interference in $B^\pm$\ra$DK^\pm$ decays, with the $D$-meson decaying to a %%@
hadronic final state. The analysis of quantum-correlated decays of the $\psi(3770)$ at CLEO-c provides invaluable information on the strong-phase %%@
difference between the $D^0$ and $\overline{D^0}$ across the Dalitz plane. Results from analyses of the decays $D$\ra$K^0\pi^+\pi^-$ and %%@
$D$\ra$K^0K^+K^-$ will be presented.
\end{abstract}
\section{Introduction}
The angle $\gamma$ is currently the least precisely known of the CKM angles of the unitarity triangle. The average value of $\gamma$ obtained from %%@
direct measurements is $70^{+27\circ}_{-29}$~\cite{CKMfitter}. A theoretically clean strategy to extract $\gamma$ is to exploit the interference %%@
between $B^\pm$\ra$D^0K^\pm$ and $B^\pm$\ra$\overline{D^0}K^\pm$ decays in which the $D^0$ and $\overline{D^0}$ decay to the same final state %%@
$F$~\cite{GGSZ}. 
This method requires precise knowledge of how the strong-phase difference between the $D^0$ and $\overline{D^0}$ varies across the Dalitz plane. %%@
Quantum-correlated data from CLEO-c provide a unique opportunity to measure this variation. This paper describes the procedure used when $F$ is either %%@
$K^0\pi^+\pi^-$ or $K^0K^+K^-$, collectively denoted $K^0h^+h^-$. The strong phase distribution for $D$\footnote{Henceforth, $D$ indicates either %%@
$D^0$ or $\overline{D^0}$}\ra$K^0\pi^+\pi^-$ is in general different to that for $D$\ra$K^0K^+K^-$, but the formalism is the same. 
\section{Determination of $\gamma$ with $B^\pm$\ra$D(K^0h^+h^-)K^\pm$}
The expression for the decay amplitude of $B^\pm$\ra$D(K^0h^+h^-)K^\pm$ is:
\begin{equation}
A(B^\pm\rightarrow D(K^0h^+h^-)K^\pm) \propto f_D(x,y) + r_Be^{i(\delta_B\pm\gamma)}f_D(y,x)
\end{equation}
where $r_B\sim 0.1$ is the ratio of the magnitudes of the amplitudes of suppressed and favoured $B^\pm$\ra$DK^\pm$ decays~\cite{CKMfitter}, $\delta_B$ %%@
is the CP-invariant strong-phase difference between interfering $B^\pm$ decays, $x$ and $y$ are the squared invariant masses $m^2(K^0h^+)$ and %%@
$m^2(K^0h^-)$ respectively and $f_D(x,y)\equiv|f_D(x,y)|e^{i\delta_D(x,y)}$ is the $D$-decay amplitude. Neglecting CP violation, the square of $A$ %%@
contains the \textit{strong-phase difference} term $\Delta\delta_D\equiv\delta_D(x,y)-\delta_D(y,x)$. In order to extract $\gamma$ from the difference %%@
in $B^\pm$ decay rates across the Dalitz plane, $\Delta\delta_D$ must be determined. 

Previous studies have modelled the $D$ decay using several intermediate two-body resonances~\cite{BaBar2008,Belle2008}. This introduces a model %%@
systematic uncertainty of $7-9^{\circ}$ to the value of $\gamma$ which is likely to be the main limitation at future b-physics experiments.

\begin{figure}[h]
\begin{minipage}[b]{10pc}
\includegraphics[width=10pc]{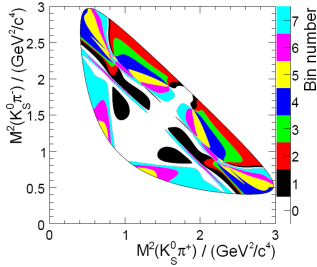}
\end{minipage}
\hspace{1.5pc}
\begin{minipage}[b]{26pc}
\caption{\label{KsPiPiSPD}The strong-phase difference for $K^0_S\pi^+\pi^-$, binned uniformly in eight bins of $\Delta\delta_D$}
\end{minipage}
\end{figure}

An alternative approach, which is the subject of the remainder of this paper, is to use external information on the strong-phase difference in a %%@
binned fit to the distribution of events across the Dalitz plane~\cite{GGSZ}. A good choice of binning is to divide the Dalitz plane into regions of %%@
similar $\Delta\delta_D$~\cite{BandP} based on a model of the $D$-decay. An incorrect model will not bias the measurement of $\gamma$, but will reduce %%@
the statistical precision. An example of a particular binning, for $K^0_S\pi^+\pi^-$, is given in Figure~\ref{KsPiPiSPD}. This uses a model from %%@
Ref.~\cite{BaBar2005}. 

The Dalitz plot is binned symmetrically about $y=x$; bins below this axis are numbered $i$ and those above $-i$. The number of events in the %%@
$i^{\scriptsize{\textrm{th}}}$ bin of the $D$\ra$K^0_Sh^+h^-$ Dalitz plot in the decay $B^\pm$\ra$D(K^0_Sh^+h^-)K^\pm$ is dependent upon $c_i$ and %%@
$s_i$ which are the average cosine and sine of the strong-phase difference across the $i^{\scriptsize{\textrm{th}}}$ bin~\cite{GGSZ}. 
\section{Determination of $\Delta\delta_D$ with quantum-correlated decays of the $\psi(3770)$} 
The $D$-mesons are quantum-correlated with an overall CP of -1, so knowledge of the CP state of one of the pair reveals the CP state of the other. %%@
When one $D$ decays to $K^0h^+h^-$, the decay product of the other $D$ is denoted the opposite-side \textit{tag}. 

Define $K_i$ as the number of events in the $i^{\scriptsize{\textrm{th}}}$ bin of the flavour-tagged $K^0_Sh^+h^-$ Dalitz plot. $c_i$ and $s_i$ can %%@
then be expressed as:
\begin{equation}
c_i\equiv\frac{a_D^2}{\sqrt{K_iK_{-i}}}\int_i|f_D(x,y)||f_D(y,x)|\cos[\Delta\delta_D(x,y)]dxdy
\end{equation}
and
\begin{equation}
s_i\equiv\frac{a_D^2}{\sqrt{K_iK_{-i}}}\int_i|f_D(x,y)||f_D(y,x)|\sin[\Delta\delta_D(x,y)]dxdy
\end{equation}
where $a_D$ is a normalization factor. Analogous quantities, denoted $K_i'$, $c_i'$ and $s_i'$, exist for $K^0_Lh^+h^-$. To first order, $c_i'=c_i$ %%@
and $s_i'=s_i$, but there are second-order differences due to doubly Cabibbo suppressed contributions to the $D$\ra$K^0_Lh^+h^-$ decay amplitude.

$c_i^{(\prime)}$ can be determined from CP-tagged $D$ decays to $K^0h^+h^-$. The number of events in the $i^{\scriptsize{\textrm{th}}}$ bin of a %%@
$K^0_{S(L)}h^+h^-$ Dalitz plot, where the opposite-side tag is a CP eigenstate, is given by: 
\begin{equation}
M_i^{(\prime)\pm}=h^{(\prime)}_{CP\pm}(K_i^{(\prime)}\pm(-1)^p\,2c_i^{(\prime)}\sqrt{K_i^{(\prime)}K_{-i}^{(\prime)}}+K_{-i}^{(\prime)})
\end{equation}
where $h^{(\prime)}_{CP\pm}$ is a normalization factor, $p=0$ for $K^0_Sh^+h^-$ and $1$ for $K^0_Lh^+h^-$.

In order to determine $s_i^{(\prime)}$, decays in which both $D$-mesons decay to $K^0h^+h^-$ are required. The number of events in the %%@
$i^{\scriptsize{\textrm{th}}}$ Dalitz plot bin of $D$\ra$K^0_Sh^+h^-$ and the $j^{\scriptsize{\textrm{th}}}$ of $D$\ra$K^0_{S(L)}h^+h^-$ is:
\begin{equation}
M^{(\prime)}_{i,j}=h^{(\prime)}_{corr}(K_iK_{-j}^{(\prime)}+K_{-i}K_j^{(\prime)} - %%@
(-1)^p\,2\sqrt{K_iK_{-j}^{(\prime)}K_{-i}K_j^{(\prime)}}(c_ic_j^{(\prime)}+s_is_j^{(\prime)}))
\end{equation}
where $h^{(\prime)}_{corr}$ is a normalization factor. 

The parameters $c_i$, $s_i$, $c_i'$ and $s_i'$ are extracted by minimizing a maximum likelihood function based on the expected and observed values of %%@
$M_i^{(\prime)}$ and $M_{i,j}^{(\prime)}$. The quantities $(c_i-c_i')$ and $(s_i-s_i')$, predicted by the model, are used as a constraint in the fit. %%@
The consequences of variations in the predicted values are evaluated when assigning systematic errors.
\section{Event Selection at CLEO-c}
An integrated luminosity of $(818\pm 8)\textrm{ pb}^{-1}$ of $\psi(3770)$\ra$D^0\overline{D^0}$ data were recorded at CLEO-c. The tags selected for %%@
$K^0K^+K^-$ and $K^0\pi^+\pi^-$ are shown in Tables~\ref{K0KKsels} and~\ref{K0PiPisels} respectively. In order to maximize statistics, a large number %%@
of tags were selected. Approximately 23,000 events were selected for $K^0\pi^+\pi^-$ and approximately 1,900 for $K^0K^+K^-$. The $\Delta\delta_D$ %%@
bins mentioned earlier were appropriately populated using selected event yields. 

$K^\pm$, $\pi^\pm$, $\pi^0$ and $\eta$ particles were selected using kinematic and reconstruction quality criteria~\cite{CLEO}. Suitable invariant %%@
mass cuts were used to select composite particles; for example, for the $\omega$\ra$\pi^+\pi^-\pi^0$ decay, the invariant mass of the %%@
$\pi^+\pi^-\pi^0$ system was constrained to lie within 20 MeV of the nominal $\omega$ mass. 

For tags containing a $K^0_L$, a different approach was used, because $\sim97\%$ of $K^0_L$ particles escape the CLEO-c detector. All other particles %%@
in the $K^0_L$ tag were reconstructed and the square of the missing mass, $m_{miss}^2$, was computed. Events could then be accepted or rejected %%@
depending on where they lay on the $m_{miss}^2$ plane. 

All raw yields were efficiency-corrected and background-subtracted. Flat backgrounds were estimated using sidebands and peaking backgrounds were %%@
estimated from Monte Carlo. The CLEO-c environment is clean; most backgrounds were low, between 1 and 10\%. 
\begin{table}[h]
\caption{\label{K0KKsels}Tags selected for $K^0K^+K^-$}
\scriptsize
\begin{center}
\begin{tabular}{p{7.5pc} p{30pc}}
\br
Tag Group & Opposite-Side Tags\\
\mr
$K^0_SK^+K^-$ vs CP+ & $K^+K^-$, $\pi^+\pi^-$, $K^0_S\pi^0\pi^0$, $K^0_L\pi^0$, $K^0_L\eta(\gamma\gamma)$, $K^0_L\omega(\pi^+\pi^-\pi^0)$, %%@
$K^0_L\eta(\pi^+\pi^-\pi^0)$, $K^0_L\eta'(\pi^+\pi^-\eta)$\\
$K^0_SK^+K^-$ vs CP- & $K^0_S\pi^0$, $K^0_S\eta(\gamma\gamma)$, $K^0_S\omega(\pi^+\pi^-\pi^0)$, $K^0_S\eta(\pi^+\pi^-\pi^0)$, %%@
$K^0_S\eta'(\pi^+\pi^-\eta)$, $K^0_L\pi^0\pi^0$\\
$K^0_SK^+K^-$ vs $K^0h^+h^-$ & $K^0_SK^+K^-$, $K^0_LK^+K^-$, $K^0_S\pi^+\pi^-$, $K^0_L\pi^+\pi^-$\\
$K^0_SK^+K^-$ vs Flavour & $K^\pm\pi^\mp$, $K^\pm\pi^\mp\pi^0$\\
$K^0_LK^+K^-$ vs CP+ & $K^+K^-$, $\pi^+\pi^-$, $K^0_S\pi^0\pi^0$\\
$K^0_LK^+K^-$ vs CP- & $K^0_S\pi^0$, $K^0_S\eta(\gamma\gamma)$, $K^0_S\omega(\pi^+\pi^-\pi^0)$, $K^0_S\eta(\pi^+\pi^-\pi^0)$, %%@
$K^0_S\eta'(\pi^+\pi^-\eta)$\\
$K^0_LK^+K^-$ vs $K^0h^+h^-$ & $K^0_S\pi^+\pi^-$\\
$K^0_LK^+K^-$ vs Flavour & $K^\pm\pi^\mp$, $K^\pm\pi^\mp\pi^0$\\
\br
\end{tabular}
\end{center}
\end{table}
\begin{table}[h]
\caption{\label{K0PiPisels}Tags selected for $K^0\pi^+\pi^-$}
\scriptsize
\begin{center}
\begin{tabular}{p{7.5pc} p{17pc}}
\br
Tag Group & Opposite-Side Tags\\
\mr
$K^0_S\pi^+\pi^-$ vs CP+ & $K^+K^-$, $\pi^+\pi^-$, $K^0_S\pi^0\pi^0$, $K^0_L\pi^0$\\
$K^0_S\pi^+\pi^-$ vs CP- & $K^0_S\pi^0$, $K^0_S\eta(\gamma\gamma)$, $K^0_S\omega(\pi^+\pi^-\pi^0)$\\
$K^0_S\pi^+\pi^-$ vs $K^0h^+h^-$ & $K^0_S\pi^+\pi^-$, $K^0_L\pi^+\pi^-$\\
$K^0_S\pi^+\pi^-$ vs Flavour & $K^\pm\pi^\mp$, $K^\pm\pi^\mp\pi^0$, $K^\pm\pi^\mp\pi^\pm\pi^\mp$, $K^\pm e^\mp\nu_e$\\
$K^0_L\pi^+\pi^-$ vs CP+ & $K^+K^-$, $\pi^+\pi^-$\\
$K^0_L\pi^+\pi^-$ vs CP- & $K^0_S\pi^0$, $K^0_S\eta(\gamma\gamma)$\\
$K^0_L\pi^+\pi^-$ vs Flavour & $K^\pm\pi^\mp$, $K^\pm\pi^\mp\pi^0$, $K^\pm\pi^\mp\pi^\pm\pi^\mp$\\
\br
\end{tabular}
\end{center}
\end{table}
\section{Results for $D$\ra$K^0\pi^+\pi^-$}
\begin{figure}[h]
\begin{minipage}[b]{10pc}
\includegraphics[width=10pc]{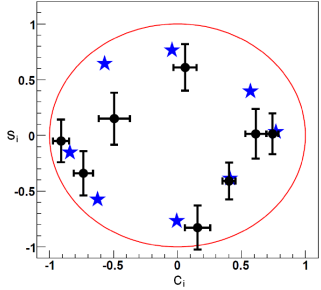}
\end{minipage}
\hspace{1.5pc}
\begin{minipage}[b]{26pc}
\caption{\label{ci_fvm}Fit results and model predictions for $c_i$ and $s_i$ for $D$\ra$K^0_S\pi^+\pi^-$. Dots indicate fit results and stars indicate %%@
model predictions.}
\end{minipage}
\end{figure}

Fit and predicted values of $c_i$ and $s_i$ for $D^0$\ra$K^0_S\pi^+\pi^-$ are shown in Figure~\ref{ci_fvm}. Systematic errors, including those taking %%@
into account variations in the model, are relatively small.

In order to understand the impact these results have on the $\gamma$ measurement, a toy Monte Carlo study of $B^\pm$\ra$DK^\pm$ was performed, with %%@
enough data so the statistical uncertainty associated with B decays was minimal. $r_B$, $\delta_B$ and $\gamma$ were fit, with initial values %%@
respectively of 0.1, $130^{\circ}$ and $60^{\circ}$. The uncertainty on $\gamma$ which propagates through from the uncertainty on $c_i$ and $s_i$ was %%@
found to be $1.7^{\circ}$~\cite{CLEO}. This is a large improvement on the current model-dependent determinations. 
\section{Preliminary Results for $D$\ra$K^0K^+K^-$}
\begin{figure}[h]
\includegraphics[width=25pc]{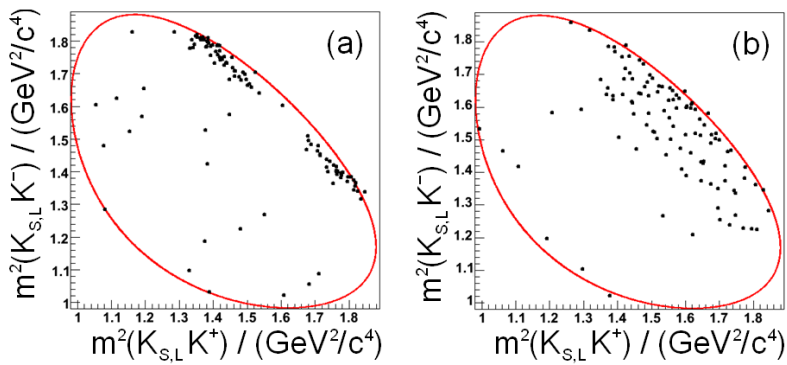}
\hspace{1pc}
\begin{minipage}[b]{11.5pc}
\caption{\label{K0KKDPs}Dalitz plots for CP-tagged $K^0K^+K^-$. \\ (a): $K^0_SK^+K^-$~vs~CP+ and $K^0_LK^+K^-$~vs~CP-. \\ (b): $K^0_SK^+K^-$~vs~CP- %%@
and $K^0_LK^+K^-$~vs~CP+. Dots indicate data points. The physical Dalitz plot boundary is delimited by a solid line.}
\end{minipage}
\end{figure}

Studies of $D$\ra$K^0K^+K^-$ are ongoing. Combined $K_{(S,L)}^0K^+K^-$ CP-tagged Dalitz plots are shown in Figure~\ref{K0KKDPs} and exhibit striking %%@
differences; this is a consequence of the entanglement of the $D^0\overline{D^0}$ system. $K^0_SK^+K^-$ against CP+ tags is in a CP- state, hence %%@
decays predominantly via $K_S\phi$. Similarly, $K^0_SK^+K^-$ against CP- tags mostly decays via $K^0_L\phi$. The $\phi$ resonance is narrow so most %%@
points lie close to $m_{K^+K^-}^2 = m_\phi^2$. This effect is not seen for the other CP-tagged data because the $K^0\phi$ resonance is not present.
\section*{References}
%\bibliography{LLWI}

\end{document}